\documentclass[aps,amssymb,prd,showpacs,twocolumn]{revtex4}
\usepackage{amsmath}
\begin{document}

\title{
Einstein boundary conditions for the 3+1 Einstein
   equations}

\author{Simonetta Frittelli}
\email{simo@mayu.physics.duq.edu}
\affiliation{Department of Physics, Duquesne University,
       Pittsburgh, PA 15282}
\author{Roberto G\'omez}
\email{gomez@psc.edu}
\affiliation{Pittsburgh Supercomputing Center, 4400 Fifth Avenue,
     Pittsburgh, PA 15213}

\date{\today}

\begin{abstract}

In the 3+1 framework of the Einstein equations for the case of
vanishing shift vector and arbitrary lapse, we calculate explicitly
the four boundary equations arising from the vanishing of the
projection of the Einstein tensor along the normal to the boundary
surface of the initial-boundary value problem. Such conditions take
the form of evolution equations along (as opposed to across) the
boundary for certain components of the extrinsic curvature and for
certain space-derivatives of the three-metric. We argue that, in
general, such boundary conditions do not follow necessarily from the
evolution equations and the initial data, but need to be imposed on
the boundary values of the fundamental variables. Using the
Einstein-Christoffel formulation, which is strongly hyperbolic, we
show how three of the boundary equations up to linear combinations
should be used to prescribe the values of some incoming
characteristic fields. Additionally, we show that the fourth one
imposes conditions on some outgoing fields.

\end{abstract}
\maketitle

The advent of sensitive gravitational wave detectors such as
LIGO~\cite{ligo} and LISA~\cite{lisa} motivates the need for
accurate templates of gravitational wave emission from powerful
sources such as the collisions of black holes with other black holes
or with neutron stars~\cite{finnthorne}.  Even though the early and
late stages of such merging events can be modelled analytically by
perturbative methods~\cite{pricepullin,pricepullinetal}, the actual
merger phase is highly non-linear and can only be expected to be
modelled accurately by numerical simulations. Most numerical
simulations of dynamical black-hole spacetimes break down well
before the relevant information can be extracted, for reasons that
are yet to be pinned down. Numerical analysis on the side, there are
many fundamental issues in the Einstein equations that may be
relevant to the stability of any numerical implementation of the
binary black hole merger, among them: intrinsic ill-posedness of the
related Cauchy problem, constraint violations, poor physical choices
of binary black hole data and poor choices of boundary conditions.
The solution to long term numerical instabilities of binary black
hole mergers is likely to involve an orchestrated control of all
these factors, in addition to the other elements of general
numerical analysis that apply to time-dependent problems with
boundaries.  In this work, we point out that even before thinking
about a numerical integration of the Einstein equations in a region
with timelike boundaries, the possible implications of the Einstein
equations to the boundaries need to be taken into consideration, as
a matter of principle.  To date, most numerical simulations with
boundaries have largely disregarded this issue, as have most
mathematical studies of the mixed initial-boundary value problem of
the Einstein equations, with exceptions explicitly noted in the
following.

Consider the 3+1 formulation of the vacuum Einstein equations with vanishing
shift vector, in which the metric has the form
\begin{equation}\label{metric}
    ds^2 = -\alpha^2 dt^2 + \gamma_{ij}dx^idx^j
\end{equation}

\noindent where $\alpha$ is the lapse function and $\gamma_{ij}$
is the Riemannian metric of the spatial three-slices of fixed value
of $t$. The Einstein equations $G_{ab}=0$ take the ADM
form~\cite{yorksources}
\begin{eqnarray}\label{adm}
    \dot{\gamma}_{ij} &=& - 2\alpha K_{ij}, \\
    \dot{K}_{ij} &=& \alpha \left(R_{ij} - 2 K_{il}K^l{}_j
            +K K_{ij}\right) -D_iD_j\alpha,
\end{eqnarray}

\noindent with the constraints
\begin{eqnarray}
    R - K_{ij}K^{ij} + K^2 &=& 0,   \\
    D_jK^j{}_i-D_i K &=& 0.
\end{eqnarray}

\noindent Here an overdot denotes a partial derivative with respect
to the time coordinate $\partial/\partial t$, indices are raised
with the inverse metric $\gamma^{ij}$, $D_i$ is the covariant
three-derivative consistent with $\gamma_{ij}$ , $R_{ij}$ is the
Ricci curvature tensor of $\gamma_{ij}$, $R$ its Ricci scalar and
$K\equiv \gamma^{ij}K_{ij}$. Because there are no time derivatives
of the lapse in any of the equations, the lapse function is
completely free.  Because a time derivative of the constraints turns
out to be a linear combination of the constraints themselves and
their space derivatives by virtue of the evolution equations, they
will vanish at subsequent times in the domain  of dependence of the
initial slice if vanishing initially (how stable the propagation of
the zero values of the constraints is to small perturbations does
not concern us at this time~\cite{frittellinote}).

Strictly speaking, the initial-value problem consists of finding a
spatially periodic or square integrable solution to the Einstein
equations in the half-spacetime $t\ge 0$.  In practice, one looks for
a spatially non-periodic solution in $t\ge 0, x^i\in {\cal V}$, where
$\cal V$ is a box in $R^3$. There arises the problem of specifying
data on the timelike boundaries of the region of interest. The problem
is rooted in the fact that in order to specify such data one needs to
know the values of the solution on the boundary for all time, which,
of course, one does not know. In an initial value problem without
constraints, the boundary values may or not be constrained by the
choice of initial data.  It seems to us that if the problem, in
addition, has constraints, the boundary values will know it, and it
should be natural to expect consequences to the boundary values from
the presence of the constraints. The point has been raised in a
precise sense by Stewart~\cite{stewart98} and further pursued in a
similar direction by Calabrese et al~\cite{calabrese01} and by
Szil\'{a}gyi and Winicour~\cite{belawell}.

The existence of at least some such consequences is intuitively
predictable from the point of view that the 3+1 equations derive
from a 4-d covariant formulation, which is more fundamental, that
is: $G_{ab}=0$. The 3+1 split is simply a choice of linearly
independent combinations of the ten components of the Einstein
tensor. The choice is precisely that which removes any time
derivatives of second order from four of the equations, and is
accomplished by projecting the Einstein tensor along the direction
normal to the slicing $G_{ab}n^b=0$.

One can easily convince oneself (or else refer to \cite{0302032})
that the projection of the Einstein tensor along the direction
normal to a surface given by a fixed value of any of the
coordinates $x^i$ removes all second derivatives with respect to
that coordinate $x^i$. Let's say we look at our boundary at a
fixed value of $x$ and write down the projection of the Einstein
tensor of the metric (\ref{metric}) along the normal vector $e^b =
(0, \gamma^{xi})$, which may be transformed to unit length by
dividing by $\sqrt{\gamma^{xx}}$, but that is clearly not
necessary. The four components of the projection of the Einstein
tensor along $e^b$ are
\begin{subequations}\label{bcs}
\begin{eqnarray}
G_{tb}e^b &=& -\frac12 \gamma^{ix}\left((\ln \gamma),_{it}
          -\gamma^{kl}\dot{\gamma}_{ik,l}\right) 
-KD^x\alpha +K^x_kD^k\alpha\nonumber\\ &&
         + \alpha\left( \gamma^{kl}\Gamma^j_{kl}K^x_j
           +\gamma^{ix}\Gamma^j_{ik}K^k_j\right)
           =0  \label{det}\\
G_{yb}e^b &=& -\frac{\dot{K}^x_y}{\alpha} + KK^x_y
          + R^x_y -\frac{1}{\alpha}
        D^xD_y\alpha=0 \label{Kxy}\\
G_{zb}e^b &=& -\frac{\dot{K}^x_z}{\alpha} + KK^x_z
          + R^x_z -\frac{1}{\alpha}
        D^xD_z\alpha =0 \label{Kxz}\\
G_{xb}e^b &=& \frac{\dot{K}-\dot{K}^x_x}{\alpha}
-\frac12 (R +K^{ij}K_{ij} +K^2) +KK^x_x\nonumber\\
&&
         +R^x_x+\frac{1}{\alpha}\left(D^jD_j\alpha- D^xD_x\alpha\right)
         =0
\label{K-Kxx}
\end{eqnarray}
\end{subequations}

\noindent  Here the time derivative of the components of the extrinsic
curvature is applied after raising an index, that is: $\dot{K}^i_j
\equiv (\gamma^{ik}K_{kj}),_t$. The reader can verify that $R^x_y$ and
$R^x_z$ do not involve second derivatives with respect to $x$ of any
of the variables and that the combination $R^x_x -\frac12 R$ doesn't
either. In the framework of the ADM equations (\ref{adm}), even
though, to our knowledge, there is no clear notion of which kind of
boundary conditions would be appropriate for a unique solution, one
could think of (\ref{det}) as an evolution equation for a particular
combination of the first space-derivatives of the metric, of
Eqs.~(\ref{Kxy}) and (\ref{Kxz}) as evolution equations for $K^x_y$
and $K^x_z$ respectively, and of Eq.~(\ref{K-Kxx}) as an evolution
equation for the combination $K^y_y+K^z_z$. Or one could freely
combine them linearly to obtain evolution equations for combinations
of these fundamental variables. In any case, these are evolution
equations along the boundary itself which can be viewed as a sort of
mixed Neumann-Dirichlet boundary conditions~\cite{roberadm}.

The situation clarifies in the case that one considers a
first-order reduction of the ADM equations. In that case, the
space-derivatives of the metric are adopted as additional
variables $q_{ijk}\equiv \gamma_{ij,k}$. This has the effect of
eliminating all appearances of $x-$derivatives of the fundamental
variables from Eqs~(\ref{bcs}).  However, a straight reduction in
this fashion is not well posed in the sense that it is not
strongly hyperbolic. If one is to consider a first-order
reduction, one is much better off adopting one that is strongly
hyperbolic in the sense that there exists a complete set of null
eigenvectors of the principal symbol of the resulting system of
equations.

In a generic strongly hyperbolic system of PDE's,  the null
eigenvectors constitute characteristic fields that propagate with
the characteristic speeds of the system, and the notion of
consistent boundary conditions can be defined unambiguously, as
follows~\cite{kreissbook}. Characteristic fields that are outgoing
at the boundary are determined by the initial values and cannot be
prescribed freely at the boundary. On the other hand,
characteristic fields that are incoming at the boundary must be
prescribed in order for there to be a unique solution, and,
furthermore, can be prescribed arbitrarily (we do not concern
ourselves with stability for the moment).

In the case that there exist constraints, it seems to us that this
notion of consistent boundary conditions may still be used but must
be adapted to the particularities of a constrained system. To wit:
if the incoming characteristic fields are to lead to a solution that
satisfies the constraints inside the region of interest, then they
must ``know'' about the constraints and, in general, may not be
prescribed as freely as in the case where there are no constraints. 
In the case of the Einstein equations, we have shown that there are
four boundary equations implied by the ten Einstein equations
themselves, which are completely equivalent to linear combinations
of the evolution equations and constraints that have no second
derivatives across the boundary. Moreover, in a sense, the
boundary equations are to the boundary surface what the constraints
are to the initial slice. To see this consider the Bianchi
identities $\nabla_aG^{ab}=0$, which take the following form in a
generic coordinate system $(t,x^i)$ adapted to spacelike slices: 
\begin{equation}
	\partial_x G^{xb} = 
	-\partial_t G^{tb}
	-\partial_y G^{yb} 
	-\partial_z G^{zb}
	- {}^{4}\Gamma^a{}_{ac}G^{cb}
	-{}^{4}\Gamma^b{}_{ac}G^{ac}
\end{equation}

\noindent where ${}^4\Gamma^c{}_{ab}$ are the Christoffel symbols
of the four-metric $g_{ab}$. The right-hand side obviously contains
no derivatives with respect to $x$ higher than second order.
Therefore, $\partial_x G^{xb}$ must be of second order in
$\partial_x$ at most, which implies that $G^{xb}$ involve, at most,
first-order derivatives with respect to $x$. Of course, the
reasoning runs as well for the $y-$derivatives, the
$z-$derivatives, and the $t-$derivatives, so that $G^{ay}$ has not
second $y-$derivatives, $G^{az}$ has no second $z-$derivatives and
$G^{tb}$ have no second time derivatives. In fact, traditionally
this reasoning has been implicitly used to argue that the
constraints $G^{tb}=0$ are the only conditions needed at the initial
slice to specify a solution to the Einstein equations via the
initial-value problem~\cite{weinberg}. 

We conclude that $G_{ab}e^b\equiv g_{ac}G^{cx}=0$ contain as much
knowledge about the solution of the Einstein equations in the
interior as the boundary values may be expected to ``know''. For
this reason, we think that imposing the boundary conditions
(\ref{bcs}) on the incoming characteristic fields is necessary in
order to evolve a solution to the evolution equations from
constrained initial data that will satisfy the four constraints at
a later time slice as well.  

To illustrate the point, we now consider the Einstein-Christoffel
formulation of the 3+1 equations~\cite{fixing}.  By defining
first-order variables as the following 18 linearly independent
combinations of the space-derivatives of the metric:
\begin{equation}\label{fkij}
f_{kij} \equiv \Gamma_{(ij)k} +
\gamma_{ki}\gamma^{lm}\Gamma_{[lj]m}
+\gamma_{kj}\gamma^{lm}\Gamma_{[li]m}
\end{equation}

\noindent where $\Gamma^k{}_{ij}$ are the Christoffel symbols of
$\gamma_{ij}$, and choosing the lapse function as $\alpha\equiv Q
\sqrt\gamma$ with $Q$ assumed arbitrarily prescribed a priori, the
3+1 equations can be put in the following equivalent
form~\cite{teukolsky}:
\begin{subequations}
\begin{widetext}
\begin{eqnarray}
\dot{\gamma}_{ij} &=& -2\alpha K_{ij}           \\
\dot{K}_{ij}+\alpha\gamma^{kl}\partial_lf_{kij} &=& \alpha \{
\gamma^{kl}(K_{kl}K_{ij} - 2K_{ki}K_{lj})
+\gamma^{kl}\gamma^{mn}(4f_{kmi}f_{[ln]j} +4f_{km[n}f_{l]ij}
-f_{ikm}f_{jln} \nonumber\\
&& +8f_{(ij)k}f_{[ln]m} +4f_{km(i}f_{j)ln} -8f_{kli}f_{mnj}
+20f_{kl(i}f_{j)mn} -13f_{ikl}f_{jmn}) \nonumber\\
&& -\partial_i\partial_j\ln Q -\partial_i\ln Q\partial_j\ln Q
+2\gamma_{ij}\gamma^{kl}\gamma^{mn}(f_{kmn}\partial_l\ln Q
-f_{kml}\partial_n\ln Q)\nonumber\\
&& +\gamma^{kl} [(2f_{(ij)k}-f_{kij})\partial_l\ln Q
+4f_{kl(i}\partial_{j)}\ln Q -3(f_{ikl}\partial_j\ln Q
+f_{jkl}\partial_i\ln Q)]\}                      \\
\dot{f}_{kij} + \alpha\partial_k K_{ij} &=& \alpha \{
\gamma^{mn}[4K_{k(i}f_{j)mn} -4f_{mn(i}K_{j)k}
+K_{ij}(2f_{mnk}-3f_{kmn})] \nonumber\\
&& +2\gamma^{mn}\gamma^{pq}[K_{mp}(\gamma_{k(i}f_{j)qn}
-2f_{qn(i}\gamma_{j)k}) +\gamma_{k(i}K_{j)m}(8f_{npq}-6f_{pqn})
\nonumber\\
&& +K_{mn}(4f_{pq(i}\gamma_{j)k}-5\gamma_{k(i}f_{j)pq})]
-K_{ij}\partial_k\ln Q \nonumber\\
&& +2\gamma^{mn}(K_{m(i}\gamma_{j)k}\partial_n\ln Q
-K_{mn}\gamma_{k(i}\partial_{j)}\ln Q)\}
\end{eqnarray}
\end{widetext}
\end{subequations}

\noindent with the constraints
\begin{subequations}
\begin{eqnarray}
{\cal C} &\equiv& \gamma^{ij}\gamma^{kl}\{
2(\partial_kf_{ijl}-\partial_if_{jkl}) +K_{ik}K_{jl}
-K_{ij}K_{kl}\nonumber\\
&& +\gamma^{mn}[f_{ikm}(5f_{jln}-6f_{ljn}) + 13f_{ikl}f_{jmn}
\nonumber\\
&& +f_{ijk}(8f_{mln}-20f_{lmn}]\} = 0   
\\    
{\cal C}_i &\equiv& \gamma^{kl}\{
\gamma^{mn}[K_{ik}(3f_{lmn}-2f_{mnl})
-K_{km}f_{iln}] \nonumber\\
&&+\partial_iK_{kl}-\partial_kK_{il}\} =0   \\
{\cal C}_{kij} &\equiv& \partial_k\gamma_{ij} -2f_{kij}
+4\gamma^{lm}(f_{lm(i}\gamma_{j)k} -\gamma_{k(i}f_{j)lm})
=0\nonumber\\
&&
\end{eqnarray}
\end{subequations}

\noindent to be imposed only on the initial data. Here ${\cal C}$
and ${\cal C}_i$ are the Hamiltonian and momentum constraints,
respectively, and ${\cal C}_{ijk}$ represent the definition of the
additional 18 first-order variables $f_{kij}$ needed to reduce the
ADM equations to full first-order form (they represent
(\ref{fkij}) inverted for $\gamma_{ij,k}$ in terms of $f_{kij}$).
No other constraints are needed to pick a solution of the Einstein
equations out of the larger set of solutions of the
Einstein-Christoffel equations. 

In this language, the boundary conditions (\ref{Kxy}-\ref{K-Kxx})
require the derivatives of the lapse $\alpha$ and the components
$R^x_y, R^x_z$ and $R^x_x-(1/2)R$ expressed in terms of $f_{kij}$,
the point being that there is a way to express their occurrence so
that no $x-$derivatives of any $f_{kij}$ appears. As they stand,
they remain as evolution equations for the components
$K^x_y,K^x_z$ and $K^y_y+K^z_z$ respectively. However, (\ref{det})
takes the following form
\begin{eqnarray}\label{f}
G_{tb}e^b &=& \dot{f}^x{}_m{}^m - \dot{f}^m{}_m{}^x + \alpha
\{2K^x_l(4f^l{}_m{}^m -3f^m{}_m{}^l) \nonumber\\
&&+K^m_lf^x{}_m{}^l - K\partial^x\alpha +
K^{xl}\partial_l\alpha\} =0 
\end{eqnarray}

\noindent where $\dot{f}^k{}_i{}^j \equiv \partial_t
(\gamma^{km}f_{mil}\gamma^{jl})$. This is an evolution equation
for a particular linear combination of the components of
$f_{kij}$.

The next step is to figure out which variables these four
equations should prescribe at the boundary.  We start by recalling
the characteristic fields of the problem with respect to the unit
vector $\xi^i \equiv \gamma^{xi}/\sqrt{\gamma^{xx}}$ which is
normal to the boundary $x=x_0$ for the region $x\le x_0$. In this
case, we have that
\begin{equation}
{}^+U^j_i \equiv K^j_i + \frac{f^x{}_i{}^j}{\sqrt{\gamma^{xx}}}
\end{equation}

\noindent are six characteristic fields that are outgoing at the
boundary, and
\begin{equation}
{}^-U^j_i \equiv K^j_i -\frac{f^x{}_i{}^j}{\sqrt{\gamma^{xx}}}
\end{equation}

\noindent are another six characteristic fields that are incoming
at the boundary.  The remaining characteristic fields are the six
components of $f^y{}_i{}^j$, the six components of $f^z{}_i{}^j$
and the six components of $\gamma_{ij}$, all 18 of which travel
upwards along the boundary (they have vanishing characteristic
speeds).

Inverting to find the fundamental fields in terms of the
characteristic fields we have
\begin{eqnarray}
    K^j_i &=& \frac{{}^+U^j_i + {}^-U^j_i}{2}     \\
    f^x{}_i{}^j &=&\frac{{}^+U^j_i - {}^-U^j_i}{2\sqrt{\gamma^{xx}}}
\end{eqnarray}

\noindent One can immediately see that the boundary conditions
(\ref{Kxy}) and (\ref{Kxz}) turn into evolution equations for
${}^-U^x_y$ and ${}^-U^x_z$ assuming the outgoing fields ${}^+U^x_y$
and ${}^+U^x_z$ are known, respectively.  In fact, these equations
yield the values of ${}^-U^x_y$ and ${}^-U^x_z$ linearly in terms of
the outgoing fields and in terms of sources known from the previous
time step. So the transverse components $a=y,z$ of the projection
$G_{ab}e^b=0$ are quite welcome and are actually necessary and
consistent with the initial-boundary value problem of the Einstein
equations.

Turning now to the longitudinal component $a=x$ of $G_{ab}e^b=0$,
which is Eq.~(\ref{K-Kxx}), we see that, because
$K-K^x_x=K^y_y+K^z_z$, the equation takes the form
\begin{equation}
({}^+\dot{U}^y_y +{}^+\dot{U}^z_z) +({}^-\dot{U}^y_y +
{}^-\dot{U}^z_z) +\ldots = 0.
\end{equation}

\noindent On the other hand, because
\begin{equation}
\dot{f}^x{}_m{}^m -\dot{f}^m{}_m{}^x = \dot{f}^x{}_y{}^y
+\dot{f}^x{}_z{}^z -\dot{f}^y{}_y{}^x -\dot{f}^z{}_z{}^x
\end{equation}

\noindent then the time component $a=t$ of $G_{ab}e^b=0$, which is
Eq.~(\ref{f}), takes the form
\begin{equation}
({}^+\dot{U}^y_y +{}^+\dot{U}^z_z) -({}^-\dot{U}^y_y 
+{}^-\dot{U}^z_z)
-\dot{f}^y{}_y{}^x -\dot{f}^z{}_z{}^x +\ldots = 0.
\end{equation}

\noindent Both equations thus involve the same incoming and outgoing
characteristic fields: ${}^-U^y_y +{}^-U^z_z$  and ${}^+U^y_y
+{}^+U^z_z$. This means that they determine \textit{both\/} the
outgoing and incoming fields ${}^+ U^y_y +{}^+ U^z_z$ and  ${}^-
U^y_y +{}^- U^z_z$. One way (by no means the only one) to make  this
point explicit is the following. Subtracting the two equations  with
appropriate factors one obtains an evolution equation for ${}^-U^y_y
+{}^-U^z_z$, which is welcome as a prescription of boundary values
that are consistent with the Einstein equations:
\begin{equation}
2\alpha G_{xb}e^b -\frac{2}{\sqrt{\gamma^{xx}}} G_{tb}e^b =
{}^-\dot{U}^y_y +{}^-\dot{U}^z_z +\dots =0,
\end{equation}

\noindent which also involves time derivatives of some of the
fields that travel along the boundary, but this is fine. In fact,
we have shown already~\cite{0302032} that at least in the case of
the Einstein-Christoffel formulation with spherical symmetry, this
resulting boundary evolution equation is being used in numerical
integration and is referred to as a constraint-preserving boundary
condition~\cite{calabrese01}.

So far, the projection of the Einstein equations in the direction
normal to the boundary is seen as providing prescriptions for the
boundary values that are necessary in the light of the
initial-boundary value problem of this strongly hyperbolic
formulation.  Any solution to the ten Einstein equations must
satisfy such boundary equations. Therefore, failure to impose such
conditions on the boundary would result in a solution to the
evolution equations with constrained initial data which would not
solve all four of the constraints at a later time slice.

There remains the question of the fourth boundary equation implied
by \textit{adding} $G_{tb}e^b=0$ and $G_{xb}e^b=0$ with
appropriate factors in order to eliminate the occurrence of the
incoming fields. This equation is linearly independent of the
other three, therefore it will not be satisfied as a consequence
of imposing the other three. One can see by inspection that this
is an evolution equation for the \textit{outgoing} field
${}^+U^y_y +{}^+U^z_z$:
\begin{equation}\label{cons}
2\alpha G_{xb}e^b +\frac{2}{\sqrt{\gamma^{xx}}} G_{tb}e^b =
{}^+\dot{U}^y_y +{}^+\dot{U}^z_z +\dots =0,
\end{equation}

\noindent which involves also the time derivatives of some of the
fields that travel along the boundary, but it does not involve the
time derivative of any incoming characteristic fields.  This
equation is not ``necessary'' in the sense that the outgoing
fields are already determined by propagation from their initial
values. However, this equation must be consistent with the
outgoing fields determined by the initial values, otherwise one
would not be solving the full set of Einstein equations.
Therefore, it is actually necessary for this equation to be
satisfied. In fact, in principle, one may simply use this equation
for those outgoing characteristic fields involved, with the
confidence that this boundary prescription is consistent with the
initial values, so long as the initial values satisfy the
constraints with sufficient accuracy. The actual numerical
implementation of this boundary condition is a subject for
subsequent study and lies beyond our current scope.

In summary, in the case of the Einstein-Christoffel formulation, the
four ``Einstein boundary conditions'' defined as $G_{ab}e^b=0$
\textit{up to linear combinations} constrain the values of three
of the six incoming characteristic fields.  The other three incoming
characteristic fields are \textit{free}, since there are no other
available conditions on the boundary. In fact, the 18 first-order
constraints ${\cal C}_{ijk}$ --which are needed in order to pick an
actual solution of the Einstein equations out of the larger set of
solutions of the Einstein-Christoffel equations-- are preserved
trivially by the evolution.  This can be seen easily by taking a
time derivative of ${\cal C}_{ijk}$ and using the evolution
equations to eliminate the occurrence of time derivatives of the
fundamental variables in the right-hand side. The reader may verify
that this leads to $\dot{\cal C}_{ijk} = F_{ijk}(C, C_i, C_{ijk})$,
that is: ordinary differential equations for the propagation of
${\cal C}_{ijk}$.  This means that the 18 first-order constraints
are trivially satisfied on the boundary by virtue of the initial
data and the evolution equations, and, consequently, they may not
prescribe conditions on the incoming fields. There thus remain three
incoming fields that can be specified in any arbitrary fashion for a
solution to the Einstein equations. They may, for instance, be
``frozen'' by setting their time derivatives to zero, as is done by
some authors~\cite{teukolsky,KTS}. 

We have thus obtained a complete set of boundary conditions for the
Einstein-Christoffel formulation from the projection of the Einstein
tensor normally to the boundary and the fact that the 18 constraints
that realize the first-order reduction are preserved by ordinary
differential equations. The incoming fields that are prescribed by
the three boundary equations \textit{carry}, in a sense, the
vanishing of $G_{ab}e^b$ into the interior, where the evolution
equations hold. Since $G_{ab}e^b$ are linear combinations of the
evolution equations and the constraints, it follows that the
constraints will be satisfied in the interior. Thus the boundary
conditions preserve the constraints.

The argument makes no use of the auxiliary system of propagation of
the constraints implied by the evolution equations. Still, the
resulting boundary conditions appear to be completely consistent
with constraint propagation in the following sense. In the
three-dimensional case that interests us here, if the evolution
equations are satisfied, the constraint functions propagate
according to a strongly hyperbolic system~\cite{KTS}. In this
auxiliary system, three characteristic constraints are incoming,
three are outgoing and the rest are ``static''.  In the context of
this auxiliary system, the three incoming constraints must be set to
zero along the boundary. One may thus infer that there will be three
conditions on the boundary values of the fundamental fields that are
necessary and sufficient to preserve the constraints. Since we have
three Einstein boundary conditions and no other conditions on the
boundary, the Einstein boundary conditions and the three conditions
that arise from constraint propagation must be equivalent in some
definite sense. We argue that the equivalence is through linear
combinations of the evolution equations. In the spherically
symmetric case we have shown that this is precisely the
case~\cite{0302032}.  Work is currently in progress to demonstrate
explicitly the equivalence of the two sets of conditions in the
three dimensional case and will appear elsewhere.  Interestingly,
since constraint propagation is at the basis of Calabrese et al's
``constraint-preserving'' boundary conditions~\cite{calabrese3d}, it
would thus appear that projecting the Einstein equations normally to
the boundary would shortcut the procedure leading to
``constraint-preserving'' boundary conditions, providing an exact
nonlinear version of the conditions and a geometric basis for them
at the same time. 

We emphasize that there is a very subtle issue of consistency
between the constrained initial data traveling towards the boundary,
and the conditions $G_{ab}e^b=0$ on the boundary that apply to the
outgoing fields. The issue turns critical in numerical simulations
because of the inaccuracies involved in satisfying the initial
constraints. Clearly, the outgoing fields propagating out from the
initial data will carry and perhaps increase the effects of the
initial errors, and there is a good chance that they will fail to
satisfy (\ref{cons}), the odds of failure increasing with time. This
issue may have large consequences for achieving long time numerical
evolution. This observation applies especially to numerical
simulations implementing \textit{black hole
excision}~\cite{excision}. As far as we know, no consistency issues
in strongly hyperbolic systems with constraints have been addressed
in the reference literature on finite-difference methods such as
\cite{kreissbook}.

Interestingly, the other constrained system of equations in physics
which is oftentimes used as a model for the Einstein equations --that
is, the Maxwell equations-- does not have  this issue. When the
Maxwell equations $\partial^aF_{ab}=J_b$ are written in terms of a
vector potential $F_{ab}=\partial_{[a}A_{b]}$, the other half of the
system $\partial_{[a}F_{bc]}=0$ is identically satisfied and the time
component $A_0=\phi$ of the vector potential may be considered
arbitrary (just as the lapse function is arbitrary insofar as the 3+1
equations are concerned). It is a straightforward exercise to put the
equations in ``ADM form'' by defining a variable $K_i\equiv
\dot{A}_i$, and to show that the projection
$e^b\partial^a\partial_{[a}A_{b]}=e^bJ_b$ for $e^b=\delta^b_x$ turns
out to be exactly one of the evolution equations. Exactly the same
thing happens in the Lorenz gauge. The underlying reason being, of
course, that there is only one constraint $C\equiv
n^b\partial^a\partial_{[a}A_{b]}-n^bJ_b$ with $n^b=\delta^b_0$ which
is preserved by the evolution by means of the trivial equation
$\dot{C}=0$.

A separate issue not treated here is whether or not the ``Einstein
boundary conditions'', that is: $G_{ab}e^b=0$ \textit{up to linear
combinations,\/}  would lead to a well-posed initial-boundary value
problem for the Einstein-Christoffel formulation in the sense that
the boundary conditions preserve the  well-posedness of the initial
value problem. In this respect, Calabrese et al~\cite{calabrese3d}
show how to obtain a subset of the constraint-preserving boundary
conditions that lead to an overall well-posed initial-boundary value
problem --including constraint propagation-- for the linearized
``generalized'' Einstein-Christoffel system. The main difference
between the Einstein-Christoffel system and its modified counterpart
(aside from the fact that the variable $f_{kij}$ is defined
differently) is that the constraints of the modified system
propagate in a symmetric hyperbolic manner, a difference that does
not affect the well-posed character, so one may expect that a subset
of well-posed boundary conditions can be found in the
Einstein-Christoffel case as well. Furthermore, the ``Einstein
boundary conditions'' clearly exist and can be written down for any
formulation of the Einstein equations, including those that are
symmetric hyperbolic and that have symmetric hyperbolic constraint
propagation, by simply expressing Eqs.~(\ref{bcs}) in terms of the
fundamental variables of the problem. There are preliminary
indications that, in such cases, linear combinations of
$G_{ab}e^b=0$ exist which are well posed in the sense
of~\cite{calabrese3d}. 

But most of the numerical relativity effort is currently driven by
formulations that are not strongly hyperbolic (see, for
instance,~\cite{BSrev,SN,laguna2,bernd,choi,miller,manuela}) in
which case sensible boundary conditions are also needed, whether or
not they are well-posed. For those formulations, the
constraint-preserving boundary conditions of Calabrese et al may not
even be defined, yet the projections $G_{ab}e^b=0$ evaluated on the
boundary are always available, are easy to write in exact non-linear
form, and provide physical boundary conditions that are clearly
necessary in order to enforce the constraints, the question of
well-posedness becoming irrelevant.  

There is the question of why should numerical simulations use these
boundary equations at all if they can do just fine by simply
``pushing the boundaries out'' --even though doing so increases the 
computational requirements considerably. In this respect, we point
out the existence of the following \textit{conjecture\/} based on
the fact, observed by many authors, that the run time of numerical
simulations is noticeably shortened by growing constraint
violations.  The conjecture, yet to be proven rigorously, is that
the run time might be extended by efficiently controlling the
constraint violations. Since we have shown that the boundaries are
intimately connected with constraint propagation, if the conjecture
turns out to be true then the ``Einstein boundary conditions'' may
have a role to play in extending the run time of numerical
simulations.

\begin{acknowledgments}
We gratefully acknowledge support by the NSF under grants No.
PHY-0070624 to Duquesne University and No. PHY-0135390 to Carnegie
Mellon University.
\end{acknowledgments}


\begin{thebibliography}{26}
\bibitem[{lis()}]{lisa}
\bibinfo{note}{LISA is the projected 
Laser Interferometer Space Antenna based
in a heliocentric orbit, \texttt{lisa.gsfc.nasa.gov}}.
\end{thebibliography}

\end{document}